\documentclass[12pt]{article}
\pdfoutput=1
\usepackage[english]{babel}
\usepackage[normalem]{ulem}
\usepackage{multirow}
\usepackage{booktabs}

\usepackage[text={17.9cm,23.7cm}]{geometry}
\usepackage{soul}
\usepackage{graphicx}
\usepackage[pdftex]{color}
\usepackage{pdfcolmk}
\usepackage{amsmath}
\usepackage{amssymb}
\DeclareMathOperator\arctanh{arctanh}
\usepackage{eurosym}
\usepackage{slashed}
\usepackage{hyperref}
\usepackage{wasysym}
\usepackage{feynmp}
\usepackage{float}
\usepackage{placeins}
\usepackage{rotating}
\usepackage{tabularx}

\begin{document}
\title{\vspace{-2cm}
{\normalsize
\flushright TUM-HEP 986/15\\}
\vspace{0.6cm}
\bf Dirac dark matter with a charged mediator: \\ a comprehensive
one-loop analysis \\ of the direct detection phenomenology\\[8mm]}

\author{Alejandro Ibarra and Sebastian Wild\\[2mm]
{\normalsize\it Physik-Department T30d, Technische Universit\"at M\"unchen,}\\[-0.05cm]
{\it\normalsize James-Franck-Stra\ss{}e, 85748 Garching, Germany}
}

\maketitle
      
\begin{abstract}
We analyze the direct detection signals of a toy model consisting of a Dirac dark matter particle which couples to one Standard Model fermion via a scalar mediator. For all scenarios, the dark matter particle scatters off nucleons via one loop-induced electromagnetic and electroweak moments, as well as via the one-loop exchange of a Higgs boson. Besides, and depending on the details of the model, the scattering can also be mediated at tree level via the exchange of the scalar mediator or at one loop via gluon-gluon interactions. We show that, for thermally produced dark matter particles, the current limits from the LUX experiment on these scenarios are remarkably strong, even for dark matter coupling only to leptons. We also discuss future prospects for XENON1T and  DARWIN and we argue that multi-ton xenon detectors will be able to probe practically the whole parameter space of the model consistent with thermal production and perturbativity. We also discuss briefly the implications of our results for the dark matter interpretation of the Galactic GeV excess.
\end{abstract}

\section{Introduction}
\label{sec:introduction}

Dark matter direct detection experiments have witnessed in the last years an impressive improvement in sensitivity, reflected in particular by the recent breaking of the zeptobarn-scale barrier in the scattering cross section by the LUX experiment~\cite{Akerib:2013tjd}. Moreover, the prospects for future improvements are bright:  in the near future the XENON1T experiment~\cite{Aprile:2012zx} will extend the reach in cross section by about one order of magnitude and, in the longer term, a multi-ton xenon detector, such as DARWIN~\cite{Baudis:2012bc}, might even reach the yoctobarn scale. 

The excellent sensitivity of experiments has allowed to rule out already some well motivated models, notably those where the dark matter is in the form of Weakly Interacting Massive Particles (WIMPs) which couple at tree level to the Standard Model via the exchange of the Z-boson, such as the sneutrino in the Minimal Supersymmetric Standard Model~\cite{Falk:1994es}. On the other hand, some other scenarios have suppressed tree-level couplings to the nucleons and can naturally evade detection at current experiments. This is the case when the dark matter interaction with the nucleus is mediated by the Higgs boson, and hence suppressed by the small Higgs coupling to the proton constituents (for example, when the dark matter is a singlet scalar~\cite{McDonald:1993ex, Cline:2013gha}, an inert $SU(2)_L$ doublet~\cite{Deshpande:1977rw,LopezHonorez:2006gr,Majumdar:2006nt} a hidden vector~\cite{Hambye:2008bq} or a singlet antisymmetric tensor~\cite{Cata:2014sta}). Alternatively, this is also the case when the dark matter only couples to the leptons or to the heavy quarks, where the tree-level couplings to the nucleons simply vanish~(as occurs, for instance,  in the radiative seesaw model \cite{Ma:2006km}). Nevertheless, even if the dark matter-nucleon tree-level coupling vanishes, such a coupling is necessarily generated via quantum effects, due to the non-vanishing gauge and Higgs coupling interactions of all the Standard Model fermions. Given the excellent sensitivity of current instruments, such quantum effects could induce observable signals at experiments. The strength of the signal is, however, very model dependent and should be investigated case by case. 

In this paper we focus on scenarios with a Dirac singlet fermion as dark matter candidate which couples to a Standard Model fermion via a scalar mediator. The model contains only three free parameters, namely the dark matter mass, the mediator mass and the Yukawa coupling strength, which can be fixed in terms of the other two parameters from requiring that the dark matter population in our Universe is generated via thermal freeze-out. In this paper we a present a comprehensive analysis of the scattering process of dark matter particles with nucleons for couplings to any Standard Model fermion. The scattering arises at tree level in the case of the first generation quarks and radiatively for other fermions, namely via the one-loop exchange of a Higgs, a Z-boson and a photon for any fermion~\cite{Agrawal:2011ze,Bai:2013iqa,Bai:2014osa,Chang:2014tea,Agrawal:2014ufa,Agrawal:2014una,Kopp:2014tsa,Yu:2014mfa,DiFranzo:2013vra}, or via a box-diagram with two external gluons for colored fermions \cite{Drees:1993bu,Hisano:2010ct,Gondolo:2013wwa}. We then calculate the scattering rate with the nucleons induced by this effective interaction, which we finally confront with the upper limit reported by LUX, as well as with the projected reach of the XENON1T and DARWIN experiments. 

The paper is organized as follows. In section \ref{sec:model_introduction} we describe the model. Then, in section \ref{sec:L_eff} we discuss the effective dark matter-nucleon interaction induced in the model, for dark matter coupling to any Standard Model fermion. In section \ref{sec:event_rate} we calculate the corresponding scattering rate with nucleons for each of these possibilities and we confront the predicted rate to the present limits from the LUX experiment and to the projected sensitivities of future experiments. Finally, in section \ref{sec:conclusions} we present our conclusions and in Appendix~\ref{sec:appendix_lux_limits}, our method to calculate the event rate in LUX.

\section{Singlet Dirac dark matter with a charged mediator}
\label{sec:model_introduction}

We consider a simple extension of the Standard Model by a Dirac fermion $\chi$, singlet under the Standard Model gauge group and which constitutes our dark matter candidate, and a charged scalar particle $\eta$, which mediates the interaction between the dark matter and a Standard Model fermion $f$. Furthermore, in order to guarantee the stability of the dark matter particle, we impose a discrete $Z_2$ symmetry under which $\chi$ and $\eta$ are odd, while all the Standard Model particles are even. The most general from for the Lagrangian reads:
\begin{align}
\mathcal{L} = \mathcal{L}_\text{SM} + \mathcal{L}_\chi + \mathcal{L}_\eta + \mathcal{L}_\text{int}^\text{fermion} + \mathcal{L}_\text{int}^\text{scalar} \,.
\end{align}
Here, $\mathcal{L}_\text{SM}$ is the Standard Model Lagrangian,  which includes a potential for the Higgs doublet $\Phi$, $V=m_1^2 \Phi^\dagger \Phi +\frac{1}{2}\lambda_1 (\Phi^\dagger \Phi)^2$, while $\mathcal{L}_\chi$ and $\mathcal{L}_\eta$ are the parts of the Lagrangian involving just the dark matter particle or the scalar mediator and which are given by:
\begin{align}
\mathcal{L}_\chi &= \bar{\chi} i \slashed{\partial} \chi - m_\chi \bar{\chi} \chi \quad \text { and}\\
\mathcal{L}_\eta &= (D_\mu \eta)^\dagger (D^\mu \eta) - m_2^2 \eta^\dagger \eta - \frac12 \lambda_2 (\eta^\dagger \eta)^2 \,,
\end{align}
where the covariant derivative $D_\mu$ depends on the gauge quantum numbers of $\eta$, which will be specified below. 

On the other hand, $\mathcal{L}_\text{int}^\text{fermion}$  describes the Yukawa interaction between the dark matter particle, the scalar mediator and the Standard Model fermion $f$, which can be either a right-handed fermion singlet $f_R \in \left\{ u_R^i, d_R^i, e_R^i \right\}$, $i=1,2,3$ being a generation index, or a left-handed fermion doublet $f_L \in \left\{ Q_L^i, L_L^i \right\}$.  Then, the interaction Lagrangian reads
\begin{align}
\label{eq:L_int_fermion_fR}
\mathcal{L}_\text{int}^\text{fermion} = -y \, \eta^\dagger \bar{\chi} f_R+ \text{h.c.} \,,
\end{align}
for coupling to  a right-handed fermion and
\begin{align}
\label{eq:L_int_fermion_fL}
\mathcal{L}_\text{int}^\text{fermion} = -y \, \eta^\dagger \bar{\chi} f_L+ \text{h.c.} = 
\begin{cases}
-y \, \eta_0^\dagger \bar{\chi} \nu_L -y \, \eta_-^\dagger \bar{\chi} e_L+ \text{h.c. } \text{ for coupling to }L_L \,, \text{ and} \\
-y \, \eta_u^\dagger \bar{\chi} u_L -y \, \eta_d^\dagger \bar{\chi} d_L+ \text{h.c. } \text{ for coupling to }Q_L
\end{cases}
\end{align}
for coupling to a left-handed fermion. Notice that specifying the fermion representation fixes the gauge quantum numbers of the scalar mediator. 

Lastly  $\mathcal{L}_\text{int}^\text{scalar}$ describes the interaction between the scalar mediator and the Standard Model Higgs boson and reads:
\begin{align}
\label{eq:L_int_scalar}
 \mathcal{L}_\text{int}^\text{scalar} = 
 \begin{cases}
 - \lambda_3 (\Phi^\dagger \Phi) (\eta^\dagger \eta) &\text{ for coupling to }f_R\,, \text{ and} \\
 - \lambda_3 (\Phi^\dagger \Phi) (\eta^\dagger \eta) - \lambda_4 (\Phi^\dagger \eta) (\eta^\dagger \Phi) &\text{ for coupling to }f_L \,.
 \end{cases}
\end{align}
These interaction terms only affect the phenomenology in a significant way when the couplings are ${\cal O}(1)$. Therefore, in what follows we will neglect these interactions and we will simply set these couplings to zero. This choice implies in particular that, for the scenario where the dark matter couples to left-handed fermions, the two components of the scalar doublet $\eta$ have a common mass, which we denote by $m_\eta$.

Provided that the coupling $y$ is large enough, the dark matter candidate $\chi$  is kept in thermal equilibrium in the Early Universe with the plasma of Standard Model particles. For large parts of the parameter space, the annihilation process which is most relevant for the equilibration is $\chi \bar{\chi} \rightarrow f \bar{f}$, where $f$ is the Standard Model fermion that couples to the dark matter particle via the interaction given in Eq.~(\ref{eq:L_int_fermion_fR}) or~(\ref{eq:L_int_fermion_fL}). The thermally averaged annihilation cross section at the time of dark matter decoupling is, for Dirac fermions, well approximated by the velocity independent part of the cross section, which reads:
\begin{align}
\label{eq:sigmav_ff}
\langle \sigma v \rangle \simeq \sigma v \big|_{v \rightarrow 0} = 
\frac{y^4 N_c}{32 \pi}\, \frac{m_\chi^2 \sqrt{1-\left( m_f/m_\chi\right)^2}}{\left( m_\chi^2+m_\eta^2-m_f^2\right)^2} \,.
\end{align}
In this expression, $N_c=3$ for coupling to quarks and $N_c=1$ for coupling to leptons. Moreover, in scenarios where the dark matter particle couples to a fermion doublet $f_L$, the total annihilation cross section is given by the sum of the annihilation cross sections into the two states of the fermion doublet. As is well known, for Dirac dark matter the total annihilation cross section for the process $\chi \bar{\chi} \rightarrow f \bar{f}$ at the time of freeze-out should take the value $\langle \sigma v \rangle_\text{thermal} \simeq 4.4  \times 10^{-26} \, \text{cm}^3/\text{s}$ in order to match the dark matter density in our Universe, $\Omega_{\rm DM} h^2=0.1198\pm 0.0015$,  as measured by the Planck satellite~\cite{Planck:2015xua}.

In this work we also consider the possibility that the scalar mediator $\eta$ is close in mass to the dark matter particle $\chi$, in which case the calculation of the dark matter abundance must be modified due to the presence of coannihilation processes. Namely, if $m_\eta \lesssim 1.2\, m_\chi$, the temperature at the time of dark matter decoupling can be large enough to keep $\eta$ also in thermal equilibrium with the plasma, which efficiently depletes the number of dark matter particles through the processes $\chi \eta \rightarrow q g$, $\eta \eta^\dagger \rightarrow g g$, and $\eta_u \eta_d^\dagger \rightarrow W^+ g$, if $\chi$ couples to quarks, and through $\chi \eta \rightarrow e^- \gamma$, $\eta \eta^\dagger \rightarrow \gamma \gamma, \gamma Z, ZZ, W^+ W^-$, or $\eta_0 \eta_-^\dagger \rightarrow W^+ \gamma$, if  $\chi$ couples to leptons. In order to fully take into account all coannihilation channels in the calculation of the relic density, we use micrOMEGAs~\cite{Belanger:2013oya}, interfaced with FeynRules~\cite{Alloul:2013bka} and CalcHEP~\cite{Belyaev:2012qa} for the numerical solution of the corresponding Boltzmann equation. 

In our numerical analysis we will focus in the well motivated scenario where the Dirac dark matter particle is produced via thermal freeze-out in the early Universe. This condition fixes one of the parameters of the model in terms of the remaining two, {\it e.g.} the Yukawa coupling $y = y_{\rm thermal}(m_\chi,m_\eta)$. Then, the parameter space of the model is spanned by only two parameters, for example, the dark matter mass $m_\chi$ and the relative difference between the dark matter mass and the mediator mass, $(m_\eta-m_\chi)/m_\chi$.

\section{Effective Lagrangian for dark matter-nucleon scattering}
\label{sec:L_eff}

\begin{figure}
\begin{center}
\includegraphics[scale=0.6]{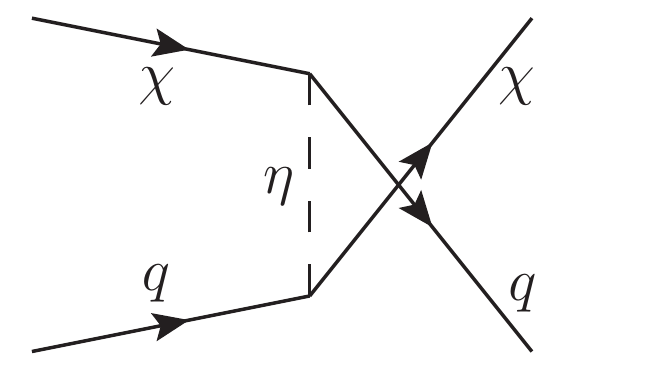}
\includegraphics[scale=0.6]{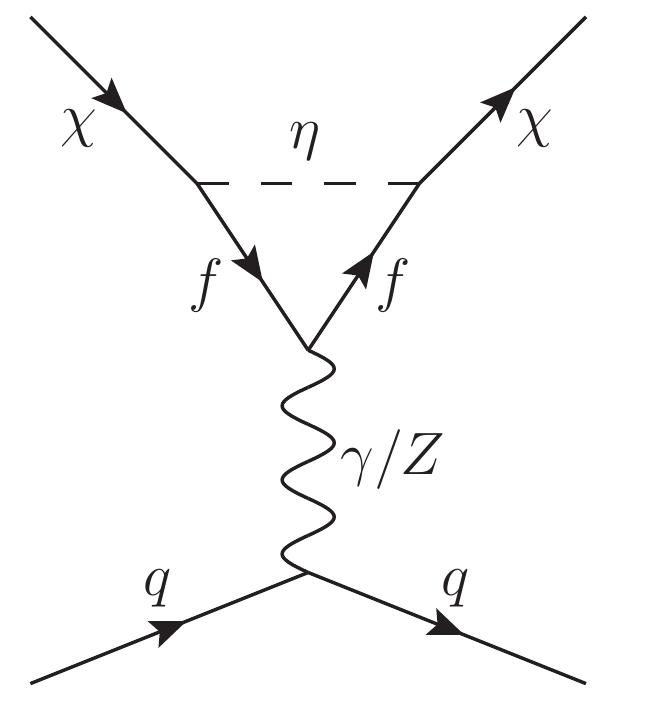}
\includegraphics[scale=0.6]{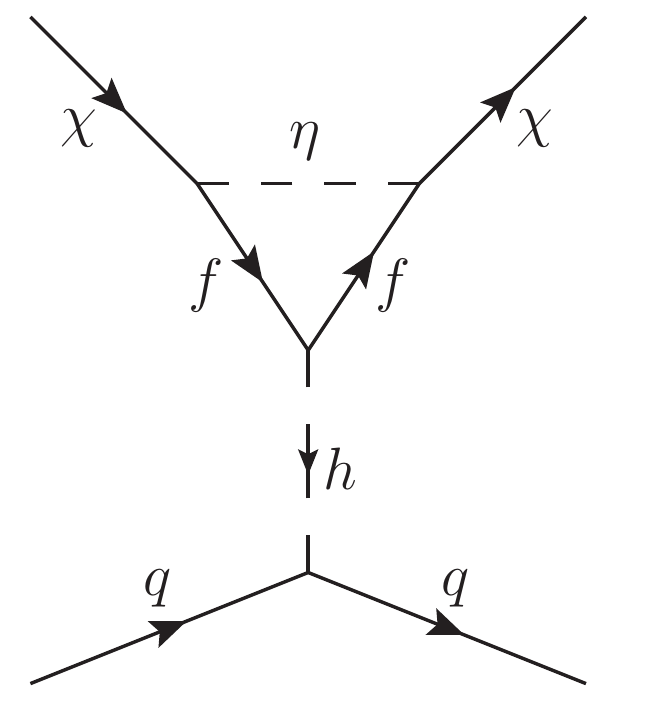}
\includegraphics[scale=0.6]{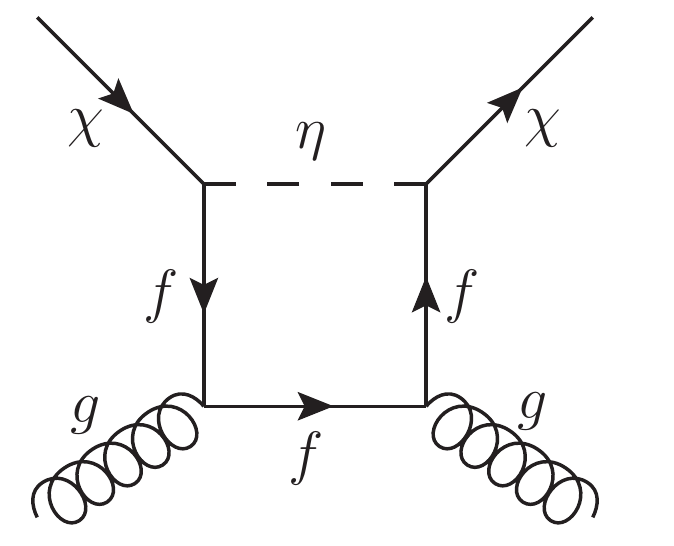}\\
(a) \hspace{3.5cm} (b) \hspace{3.5cm} (c) \hspace{3.5cm} (d)
\end{center}
\caption{\small Exemplary Feynman diagrams contributing to the dark matter-nucleon scattering: (a) tree-level scattering, (b) one-loop exchange of a photon or $Z$  boson, (c) one-loop exchange of a Higgs boson, and (d) scattering off gluons at one loop.}
\label{fig:scattering_diagrams}
\end{figure}

The interactions of a Dirac dark matter particle with a nucleon can be described  by the Feynman diagrams shown, up to one-loop in perturbation theory, in Fig.~\ref{fig:scattering_diagrams}, where we include for conciseness only one representative diagram of each class.  Panel (a) corresponds to the tree-level interaction mediated by the exchange of the scalar $\eta$. Panels (b) and (c) show penguin diagrams mediated by the photon (or the $Z$-boson) and by the Higgs boson, respectively, which arise at the one loop level in all models of Dirac dark matter with a charged scalar mediator, regardless of the choice of the Standard Model fermion. Finally, panel (d) shows a box-diagram with two external gluons, which arises also at the one loop level when the dark matter couples to a colored scalar mediator. Let us discuss separately the effective Lagrangian that arises in each case.

If the dark matter couples at tree level to a first generation quark, the dominant contribution to the scattering cross section is the tree level process $\chi q\rightarrow \chi q$ with the $t$-channel exchange of the scalar mediator $\eta$, depicted in Fig.~\ref{fig:scattering_diagrams}, panel (a). Analogously, the dark matter antiparticle $\bar\chi$  scatters off quarks, with the scalar being in this case exchanged in the $s$-channel. After a Fierz rearrangement of the corresponding matrix elements, these diagrams give rise to a vector interaction of $\chi$ with the quark, which in turn translates into an effective vector interaction of the dark matter particle with the nucleon of the form:
 \begin{align}
 \label{eq:Leff_tree}
 \mathcal{L}_\text{eff,tree} &= f_{V,\text{tree}}^{(N)}\,\bar{\chi} \gamma^\mu \chi \, \bar{N} \gamma_\mu N
 \end{align}
and which leads, in the non-relativistic limit, to a spin-independent interaction. Here, the effective couplings for protons and neutrons, $f_{V,\text{tree}}^{(p)}$  and $f_{V,\text{tree}}^{(n)}$, read
\begin{align}
\label{eq:fV_tree}
 f^{(p)}_{V,\text{tree}}&= \begin{cases} 2 \, \lambda_V &\text{for coupling to }u_R \\ \lambda_V &\text{for coupling to } d_R\\ 3 \, \lambda_V &\text{for coupling to }(u_L, d_L)\end{cases}
 \quad \text{and} \quad
  f^{(n)}_{V,\text{tree}}&= \begin{cases} \lambda_V &\text{for coupling to }u_R \\ 2 \, \lambda_V &\text{for coupling to } d_R\\ 3 \, \lambda_V &\text{for coupling to }(u_L, d_L)\end{cases}
 \end{align}
 with $\lambda_V = y^2/\left[ 8 \left( m_\eta^2- m_\chi^2\right) \right]$. Due to vector-current conservation, the spin-independent tree-level contribution is exactly zero for coupling to second- and third-generation quarks, and is of course also zero for dark matter coupling solely to a lepton. The same diagrams also give rise to an axial vector interaction of the form $f_{A,\text{tree}}^{(N)} \chi \gamma^\mu \gamma^5 \chi \bar{q} \gamma_\mu \gamma_5 q$, which leads to a spin-dependent interaction.  The coupling constants of the axial vector and the vector interaction turn out to be comparable in size in this model, therefore, given that the experimental sensitivity to the spin-independent interaction is orders of magnitude better than to the spin-dependent one, this contribution can be in most instances safely neglected. Nevertheless, it will also be included in our numerical analysis for completeness. 
 
On the other hand, and regardless of the choice of the Standard Model fermion, the Yukawa interactions Eq.~(\ref{eq:L_int_fermion_fR}) and Eq.~(\ref{eq:L_int_fermion_fL}) necessarily induce at the quantum level the scattering of dark matter particles with nucleons, due to the penguin diagrams depicted in panels (b) and (c) of Fig.~\ref{fig:scattering_diagrams}. The first diagram induces electromagnetic moments for the dark matter particle, $\chi$, the most relevant ones for Dirac dark matter being the magnetic dipole moment $\mu_\chi$ and the charge radius $b_\chi$, defined by the effective dark matter-photon Lagrangian:\footnote{This diagram also leads in addition to an anapole moment, which corresponds to the effective operator $\bar{\chi} \gamma^\mu \gamma^5 \chi \partial^\nu F_{\mu \nu}$. This operator is suppressed in the non-relativistic limit by the square of the dark matter velocity and is always subdominant in this model. This contribution, however, is also included in our numerical analysis for completeness.}
\begin{align}
 \label{eq:L_eff_gamma}
 \mathcal{L}_{\text{eff},\gamma} = \frac{\mu_\chi}{2} \bar{\chi} \sigma^{\mu \nu} \chi F_{\mu \nu} + b_\chi \bar{\chi} \gamma^\mu \chi \partial^\nu F_{\mu \nu} \,.
 \end{align}
The electromagnetic moments $\mu_\chi$ and $b_\chi$ are obtained by matching the coefficients of this effective Lagrangian to the results of an explicit calculation of the relevant loop diagrams. For dark matter coupling to a right-handed Standard Model fermion $f_R$,  and taking the limit where the transferred momentum in the dark matter-nucleus interaction goes to zero, the result reads:\footnote{We have used FeynCalc~\cite{Mertig:1990an} for parts of the computation.}
\begin{align}
\label{eq:mu_chi}
\mu_\chi =& \frac{-Q_f e N_c y^2}{32 \pi^2 m_\chi}\, \bigg[ \frac{-\Delta+1-\mu-\epsilon}{\Delta^{1/2}} \arctanh \left(  \frac{\Delta^{1/2}}{\mu+\epsilon-1}\right) + \frac12 (\epsilon-\mu) \log \left( \frac{\epsilon}{\mu} \right) -1 \bigg]\;,\\ \nonumber
b_\chi =& \frac{-Q_f e N_c y^2}{384 \pi^2 m_\chi^2}\, \bigg[ \frac{2}{\Delta^{3/2}} \left( 8 \Delta^2+ (9 \mu+7 \epsilon-5) \Delta - 4 \epsilon (3 \mu+\epsilon- 1) \right) \arctanh\left(  \frac{\Delta^{1/2}}{\mu+\epsilon-1}\right) \\
\label{eq:b_chi}
& + (8 \mu-8 \epsilon + 1) \log \left( \frac{\epsilon}{\mu} \right) + 4 \left( 4+ \frac{\mu+3 \epsilon - 1}{\Delta} \right) \bigg]\;.
\end{align}
Here, $Q_f$ is the charge of $f_R$ in units of e (such that e.g. $Q_f=-1$ for coupling to $e_R$), $N_c=3 \, (1)$ for coupling to quarks (leptons) and
\begin{align}
\Delta \equiv \mu^2 + (\epsilon-1)^2-2 \mu (\epsilon+1)\;,
\label{eq:definition_Delta}
\end{align}
with $\mu \equiv m_\eta^2/m_\chi^2$ and $\epsilon \equiv m_f^2/m_\chi^2$ (see also \cite{Chang:2014tea}). For dark matter coupling to a left-handed Standard Model fermion doublet, $\mu_\chi$ and $b_\chi$ are simply given by the sum of the corresponding expressions for both components in the doublet. 

The expression for $b_\chi$ presents, when $m_f\ll m_\eta$, a infrared divergence  $\propto \log \left( m_f^2/m_\eta^2\right)$, which enhances the scattering rate for light fermions. Note, however, that Eq.~(\ref{eq:b_chi}) was derived in the limit where the transferred momentum goes to zero and is therefore not valid in dark matter scenarios where the fermion mass constitutes the smallest energy scale in the scattering process. More specifically, when the dark matter scatters off xenon nuclei, the typical transferred momentum is $\sim 50$ MeV, therefore  for couplings to electrons and first generation quarks Eq.~(\ref{eq:b_chi}) is not valid. For these cases, and in order to cut-off the divergence, we replace in our numerical analysis the fermion mass by 50 MeV. On the other hand, and in contrast to the charge radius, the magnetic dipole moment $\mu_\chi$ has a finite value for $m_f\ll m_\eta$. 

Analogously, a dark matter interaction with the $Z$ boson arises at the one-loop level, leading to an effective vector interaction of the form: 
\begin{align}
\label{eq:Leff_Z}
 \mathcal{L}_{\text{eff},Z} = f_{V,Z}^{(N)} \,\bar{\chi} \gamma^\mu \chi \, \bar{N} \gamma_\mu N \,,
 \end{align}
 where
 \begin{align}
 f_{V,Z}^{(p)} = \left( 4 s_W^2-1 \right) \frac{G_F a_Z}{\sqrt{2}} \quad , \quad f_{V,Z}^{(n)} = \frac{G_F a_Z}{\sqrt{2}} \,.
 \label{eq:f_VZ}
 \end{align}
Here, $a_Z$ is an effective form factor which, for dark matter coupling to a right-handed fermion, is given by
\begin{align}
\label{eq:aZ_fR}
a_Z^{\left(f_R\right)} = \frac{T_3^f N_c y^2\epsilon}{16 \pi^2}\,  \bigg[ \frac12 \log \left( \frac{\epsilon}{\mu}\right) +\frac{1+\mu-\epsilon}{\Delta^{1/2}} \arctanh \left( \frac{\Delta^{1/2}}{\mu+\epsilon-1}  \right) \bigg]\;,
\end{align}
with $\Delta$ defined as in Eq.~(\ref{eq:definition_Delta}). For the case of dark matter coupling to a left-handed fermion doublet one has $a_Z^{\left(u_L, d_L\right)} = -a_Z^{\left(u_R\right)}-a_Z^{\left(d_R\right)}$ and $a_Z^{\left(\nu_L, e_L\right)} = -a_Z^{\left(e_R\right)}$. Notice that the dark matter effective coupling to the $Z$ boson scales as $(m_f/m_\chi)^2$, making this contribution subdominant in comparison to the photon exchange, except for the interesting scenario where the dark matter couples to $t_R$ or $(t_L, b_L)$.

The penguin diagram shown in panel (c) of Fig.~\ref{fig:scattering_diagrams} induces a coupling of $\chi$ to the Standard Model Higgs boson $h$, which in turn induces the effective dark matter-nucleon interaction
 \begin{align}
 \label{eq:Leff_higgs}
 \mathcal{L}_\text{eff,Higgs} &= f_{\text{S,Higgs}}^{(N)}\,\bar{\chi} \chi \, \bar{N} N\;,
 \end{align}
where
\begin{align}
\nonumber
f_{\text{S,Higgs}}^{(N)} =& \frac{-\sqrt{2} G_F m_\chi m_N f_N^{(0)}}{m_h^2} \,  \frac{3 \epsilon}{32 \pi^2 \Delta^{1/2}}   \\
&\left[ 2 \left( \epsilon^2 + \mu \left(\mu-1 \right)-\epsilon \left(1 + 2\mu \right) \right) \arctanh \left( \frac{\Delta^{1/2}}{\mu+\epsilon-1} \right) +\Delta^{1/2} \left( 2+ \left( \mu-\epsilon \right) \log\left( \frac{\epsilon}{\mu} \right) \right) \right]\;,
\end{align}
with the Higgs-nucleon coupling $f_N^{(0)} \simeq 0.345$~ \cite{Cline:2013gha}.\footnote{The effective dark matter interactions to the Higgs and to the Z-boson have also been presented in \cite{Kumar:2013hfa} in the limit $m_\chi \rightarrow 0$ for a model where the dark matter couples to $t_R$. Our results in that limit agree with the expressions given in that work.} As the coupling of the fermion $f$ to the Higgs is proportional to $m_f$, this contribution is clearly subdominant for dark matter coupling to any of the light Standard Model fermions. In the case of dark matter coupling to a third-generation quark, we find that the contribution from the $Z$ exchange is always much larger, rendering the Higgs exchange diagram subdominant for all possible cases, although it is also included in our numerical results. An additional contribution to the effective dark matter-Higgs interaction arises when the quartic couplings $\lambda_3$ and $\lambda_4$ in Eq.~(\ref{eq:L_int_scalar}) are non-vanishing. However, we find numerically that,  even for quartic couplings  $\mathcal{O}(1)$,  the Higgs interaction gives a subdominant contribution to the dark matter scattering off nucleons.
 
Finally, if the dark matter couples to quarks, it can scatter off gluons via one-loop box diagrams,  such as the one shown in panel (d) of Fig.~\ref{fig:scattering_diagrams}. This generates an effective scalar interaction of $\chi$ with nucleons, which reads
 \begin{align}
 \label{eq:Leff_gluon}
 \mathcal{L}_{\text{eff,gluon}} = f_{S,\text{gluon}}^{(N)} \, \bar{\chi} \chi \bar{N} N \,.
 \end{align} 
The effective coupling $f_{S,\text{gluon}}^{(N)}$ has been calculated for the case of Majorana dark matter interacting with a colored mediator in~\cite{Drees:1993bu,Hisano:2010ct,Gondolo:2013wwa}. By explicitly writing down the amplitudes for all relevant diagrams, it is straightforward to check that $f_{S,\text{gluon}}^{(N)}$ is exactly the same for Majorana and for Dirac dark matter. Hence, we do not repeat here the rather cumbersome expression for the effective scalar coupling, and we refer instead to these works. Also, following the detailed discussion in~\cite{Gondolo:2013wwa}, we include the gluon contribution to the effective dark matter-nucleon Lagrangian only for the case where dark matter couples to a heavy quark, {\it i.e.} $c$, $b$ or $t$. 

\section{Event rate for dark matter-nucleus scattering and experimental limits}
\label{sec:event_rate}
The effective Lagrangians given in Eqs.~(\ref{eq:Leff_tree},~\ref{eq:Leff_Z},~\ref{eq:Leff_higgs},~\ref{eq:Leff_gluon}) describing the interaction of dark matter with nucleons, as well as the effective dark matter-photon interaction defined by Eq.~(\ref{eq:L_eff_gamma}), induce the elastic scattering of dark matter off nuclei. Taking into account all these contributions, the differential scattering cross section of dark matter off a nucleus with mass $m_A$ and spin $J_A$ which recoils with energy $E_R$, reads, including only the leading terms in an expansion in $1/E_R$ and $1/v^2$:
\begin{align}
\label{eq:scattering_crosssection}
 \frac{\text{d}\sigma}{\text{d}E_R} = \, &\alpha_{\text{em}} \mu_\chi^2 Z^2 \left( \frac{1}{E_R} -\frac{m_A}{2 \mu_\text{red}^2 v^2} \right) \left[F_{\text{SI}} \left(E_R\right)\right]^2 \\\nonumber
 &+ \frac{\mu_A^2 \mu_\chi^2 m_A}{\pi v^2} \,\frac{J_A+1}{3J_A}\, \left[F_{\text{dipole}} \left(E_R\right)\right]^2 \\\nonumber
 &+ \frac{m_A}{2 \pi v^2} \left( f^{(A)}\right)^2 \left[F_{\text{SI}} \left(E_R\right)\right]^2 \,,
\end{align}
where $f^{(A)}$ is the effective scalar dark matter-nucleus coupling, defined as
\begin{align}
f^{(A)} = Z \left( f_{S,\text{gluon}}^{(p)} + f_{V,\text{tree}}^{(p)} + f_{V,Z}^{(p)} - e b_\chi - \frac{e \mu_\chi}{2 m_\chi} \right) + (A-Z) \left( f_{S,\text{gluon}}^{(n)} + f_{V,\text{tree}}^{(n)} + f_{V,Z}^{(n)}\right) \,.
\end{align}
For $F_{\text{SI}} \left( E_R \right)$, we use the Helm form factor given in~\cite{Lewin:1995rx}, while for the form factor $F_{\text{dipole}} \left(E_R\right)$ associated to the dipole-dipole scattering we follow~\cite{Banks:2010eh}. 

Finally, we translate the scattering cross section into an event rate in a xenon experiment and we compare the predicted rate with the current limit by the LUX experiment~\cite{Akerib:2013tjd}, which provides the strongest sensitivity for the model discussed in this paper. It is important to remark that one can not simply map the limits on $\sigma_\text{SI}^p$ as they are published by the corresponding collaborations onto limits on the parameter space of the model considered in this work, due to basically two reasons: first, the scattering diagrams involving the exchange of a photon only induce a dark matter coupling to protons, but not to neutrons, leading to a (potentially large) source of isospin violation in the scattering. Secondly, due to the long-range force induced by the magnetic dipole moment of dark matter, the differential scattering cross section given in Eq.~(\ref{eq:scattering_crosssection}) has a different dependence on the recoil energy $E_R$ compared to the case of standard spin-independent or spin-dependent interactions. In our numerical analysis, we derive the corresponding upper limits arising from the null-result of LUX, fully taking into account the isospin violation as well as the non-standard dependence of the differential scattering cross section with $E_R$. We provide the details of our method of calculating the event rate in LUX in Appendix~\ref{sec:appendix_lux_limits}.

We will also calculate the regions of the parameter space which can be probed by the future experiments XENON1T~\cite{Aprile:2012zx}, already under construction since 2013, and DARWIN, a design study for a next generation
multi-ton dark matter detector~\cite{Baudis:2012bc}. Following~\cite{Cushman:2013zza}, we assume that XENON1T will have a reach in  $\sigma_p^{\text{SI}}$ a factor of 40 better than the present LUX limit for dark matter masses above $\simeq 100$ GeV. For the model discussed in this work, this translates into a rescaling of the upper limits on the Yukawa coupling $y$ arising from the LUX data by a factor of $40^{1/4} \simeq 2.5$. Here, we refrain from deriving prospects for XENON1T for small dark matter masses, i.e. $m_\chi \lesssim 30$ GeV, as these will depend on so far unknown details of the detector performance near threshold. For DARWIN, we consider the most optimistic scenario where the reach in cross section reaches $\simeq 10^{-48} \text{ cm}^2$~\cite{Baudis:2012bc}, thereby approaching the irreducible background of coherent neutrino-nucleus scattering, and we derive prospects by rescaling our limits on the Yukawa coupling obtained from the LUX data by a factor of $1800^{1/4} \simeq 6.5$. As for XENON1T, we do not consider prospects for low-mass dark matter. For simplicity we also assume that DARWIN will consist only of xenon, although currently both liquid argon and xenon are discussed as possible target materials.

\begin{figure}[t]
\hspace{-0.7cm}
\includegraphics[scale=1.05]{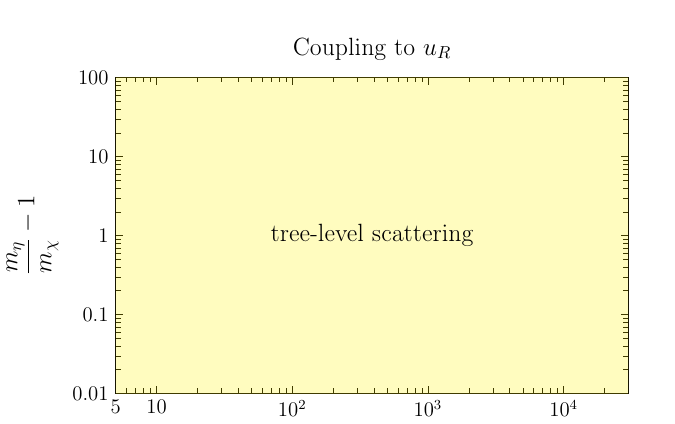}
\hspace{-1.0cm}
\includegraphics[scale=1.05]
{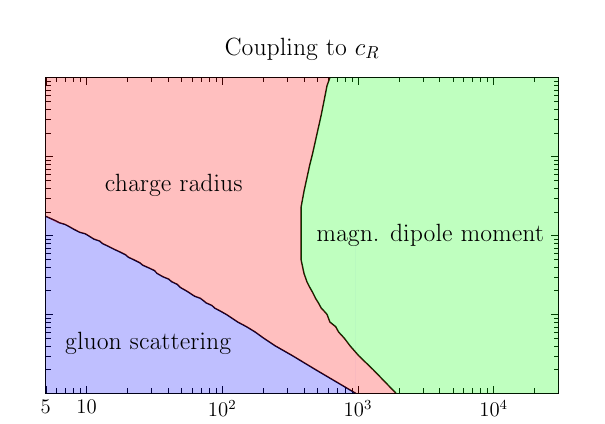}
\hspace{-1.0cm}
\includegraphics[scale=1.05]{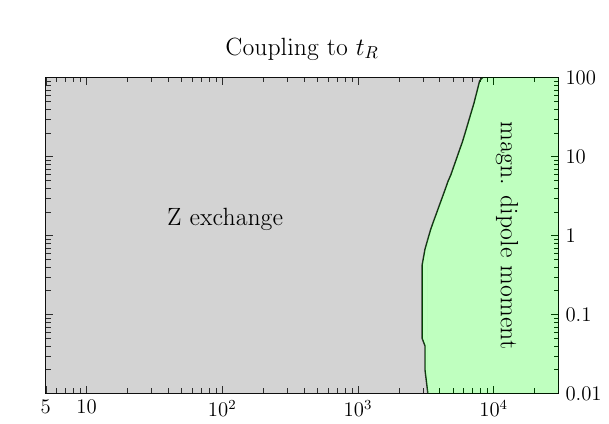}

\hspace{-0.7cm}
\includegraphics[scale=1.05]{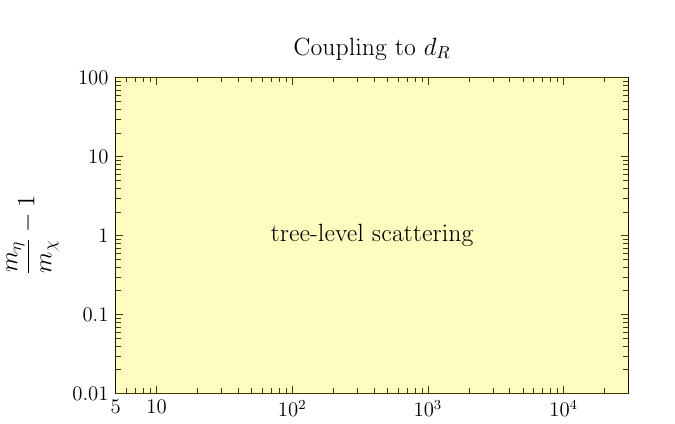}
\hspace{-1.0cm}
\includegraphics[scale=1.05]
{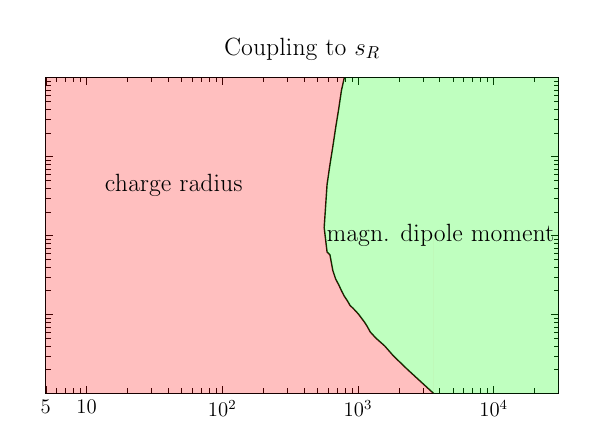}
\hspace{-1.0cm}
\includegraphics[scale=1.05]{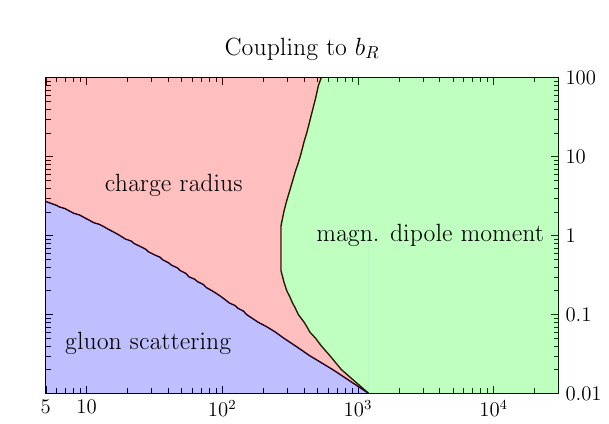}

\hspace{-0.7cm}
\includegraphics[scale=1.05]{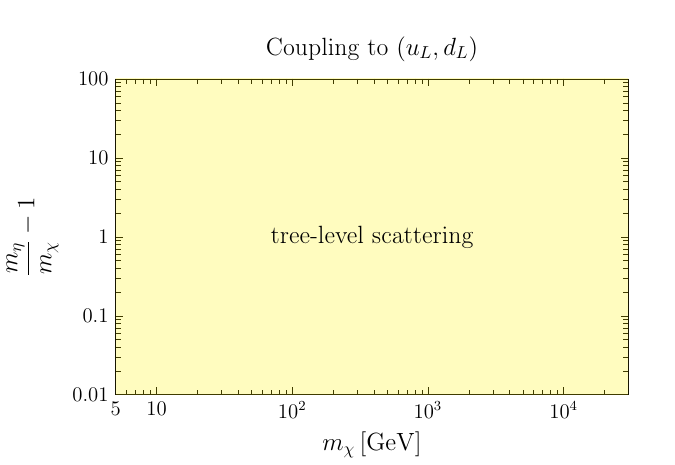}
\hspace{-1.0cm}
\includegraphics[scale=1.05]
{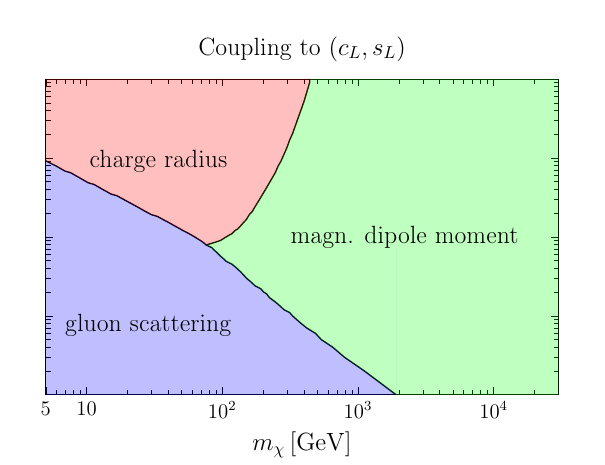}
\hspace{-1.0cm}
\includegraphics[scale=1.05]{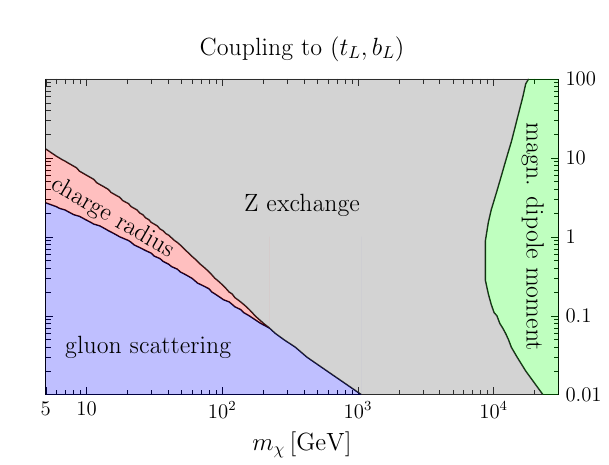}
\caption{\small Dominant contributions to the dark matter scattering rate in LUX, for coupling to right-handed up-type quarks (upper panels), right-handed down-type quarks (middle panels), and left-handed quark doublets (lower panels).}
\label{fig:MostImportantContris_QuarkCoupling}
\end{figure}

\begin{figure}
\hspace{-0.7cm}
\includegraphics[scale=1.05]{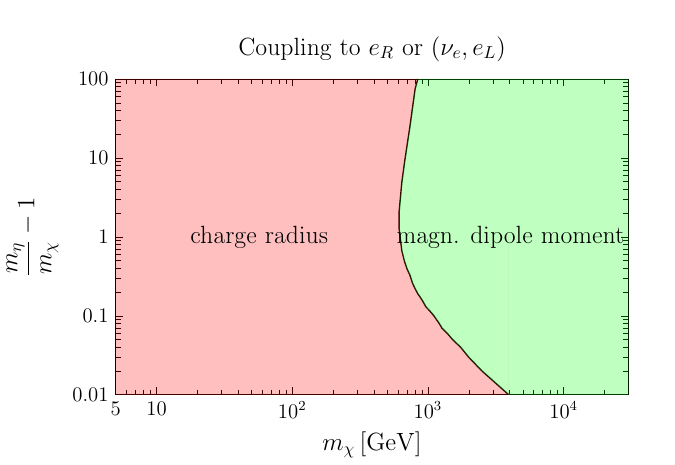}
\hspace{-1.0cm}
\includegraphics[scale=1.05]
{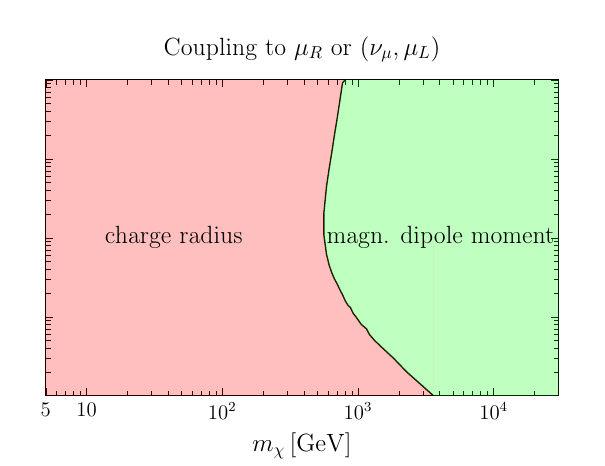}
\hspace{-1.0cm}
\includegraphics[scale=1.05]{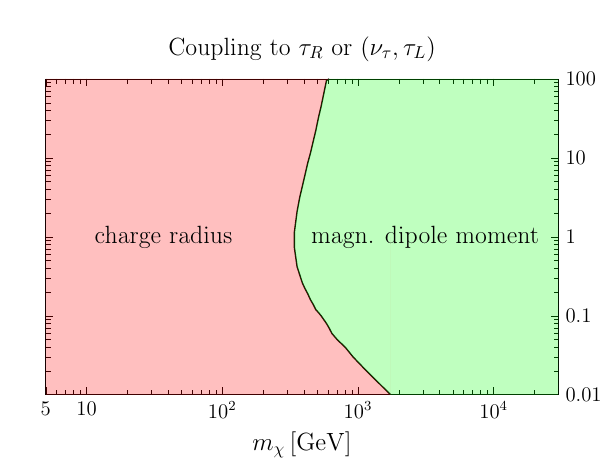}
\caption{\small Dominant contributions to the dark matter scattering rate in LUX, for coupling to leptons.}
\label{fig:MostImportantContris_LeptonCoupling}
\end{figure}

Before presenting the actual limits on the model, let us first discuss the relative importance of the different contributions to the scattering rate, which were introduced in Section~\ref{sec:L_eff} and were schematically shown in Fig.~\ref{fig:scattering_diagrams}. Which of these processes gives the largest contribution to the event rate in a direct detection experiment depends on which Standard Model fermion the dark matter particle couples to, as well as on the dark matter and mediator masses. In Figs.~\ref{fig:MostImportantContris_QuarkCoupling} and~\ref{fig:MostImportantContris_LeptonCoupling} we show, separately for each coupling scheme, which processes dominate the scattering in LUX, for given dark matter and mediator masses, $m_\chi$ and $m_\eta$. The left panels of Fig.~\ref{fig:MostImportantContris_QuarkCoupling} correspond to a scenario where the dark matter particle couples to right-handed up-quarks (top left), right-handed down-quarks (central left), or to the first generation of left-handed quarks (bottom left). As expected, in these cases the tree-level scattering of dark matter off the nucleons dominate the event rate, rendering all one-loop processes irrelevant. On the other hand, as discussed in Section~\ref{sec:L_eff}, the tree-level contribution is negligible if dark matter couples to quarks of the second or third generation, or to leptons. In all these scenarios, the one-loop processes dominate the event rate in LUX (and other direct detection experiments), as shown in Figs.~\ref{fig:MostImportantContris_QuarkCoupling} and~\ref{fig:MostImportantContris_LeptonCoupling}.  More precisely,  the scattering via photon exchange, which leads to the charge radius and magnetic dipole moment operators, is the dominant contribution to the cross section for large parts of the parameter space of each scenario. In particular, the charge radius operator typically leads to the largest event rate for  $m_\chi \lesssim 1$ TeV, while the magnetic dipole moment dominates for $m_\chi \gtrsim 1$ TeV, although the concrete values depend on the coupling scheme and on the mass splitting between the mediator $\eta$ and the dark matter particle $\chi$. This behavior can be qualitatively understood from the dimensionality of each effective operator describing the coupling of dark matter to photons (see Eq.~(\ref{eq:L_eff_gamma})), and which is dimension five for the case of magnetic dipole interactions, while it is dimension six for the charge radius. Hence, the corresponding coefficients of these operators scale differently with the dark matter mass, namely as $\mu_\chi \sim m_\chi^{-1}$ and $b_\chi \sim m_\chi^{-2}$, respectively.

On the other hand, if the dark matter particle couples to the top-quark, the scattering via the exchange of a $Z$ boson becomes relevant, and  generically dominates over the photon exchange for masses below $\simeq 5 - 10$ TeV, as shown in the right upper and right bottom panels of Fig.~\ref{fig:MostImportantContris_QuarkCoupling}. As mentioned in Section~\ref{sec:L_eff}, this is due to the dependence of the effective coupling $a_Z$ with the mass of the fermion coupling to $\chi$: from Eq.~(\ref{eq:aZ_fR}) it follows that for $m_f \ll m_\chi$ this effective coupling scales as $a_Z \propto (m_f/m_\chi)^2$, making it relevant only for the case of $\chi$ coupling to $t_R$ or $(b_L, t_L)$. Finally, we find that the dark matter-nucleon scattering induced by the one-loop coupling of dark matter to gluons is only relevant for rather small dark matter masses and small mass splittings $m_\eta/m_\chi$, for dark matter coupling to a heavy quark, as illustrated in the corresponding panels in Fig.~\ref{fig:MostImportantContris_QuarkCoupling}.

\begin{figure}[t!]
\hspace{-0.7cm}
\includegraphics[scale=1.05]{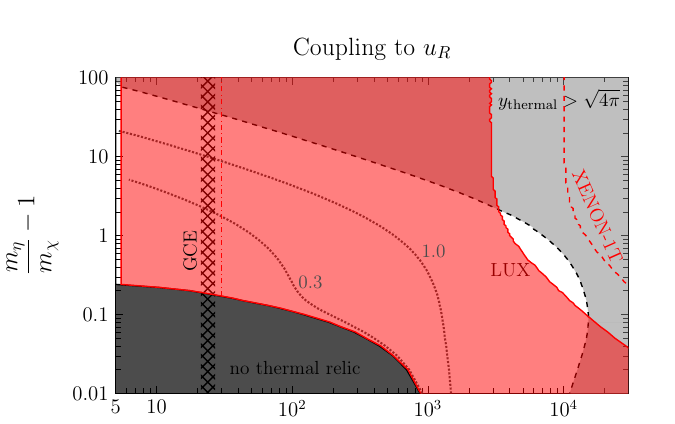}
\hspace{-1.0cm}
\includegraphics[scale=1.05]
{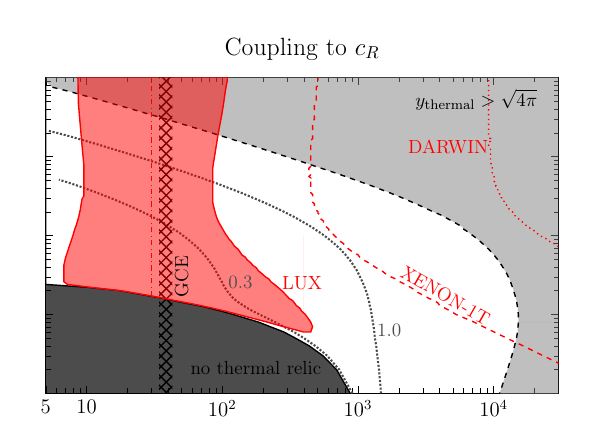}
\hspace{-1.0cm}
\includegraphics[scale=1.05]{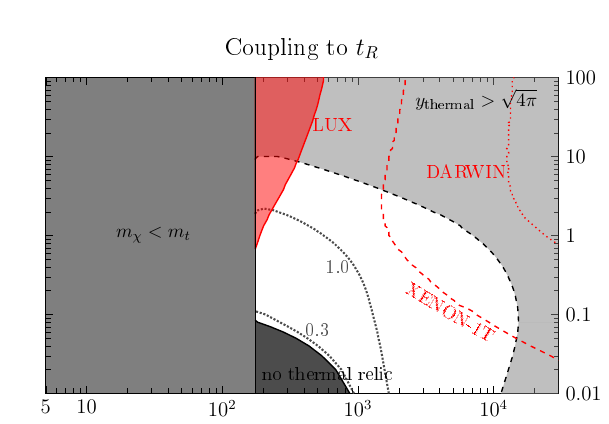}

\hspace{-0.7cm}
\includegraphics[scale=1.05]{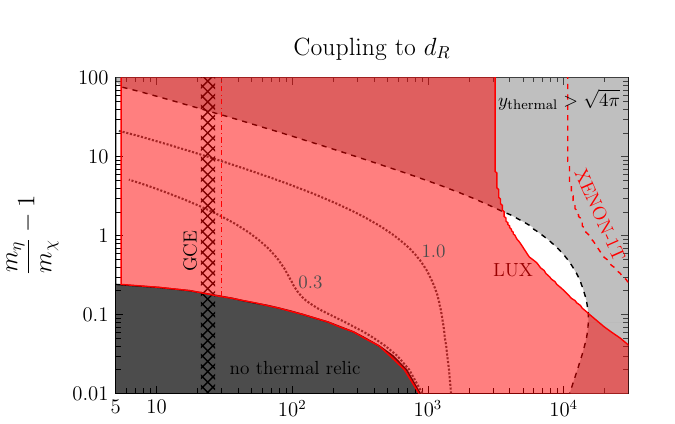}
\hspace{-1.0cm}
\includegraphics[scale=1.05]
{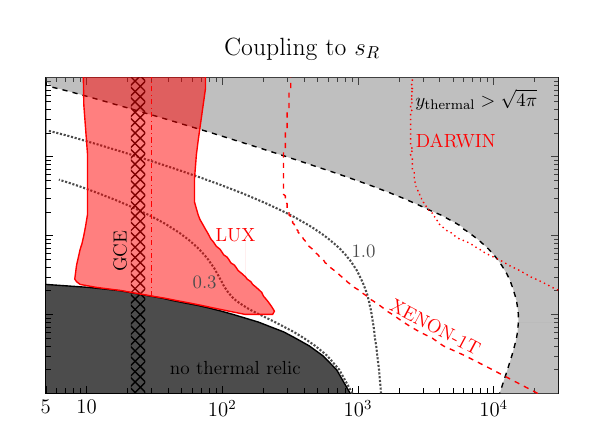}
\hspace{-1.0cm}
\includegraphics[scale=1.05]{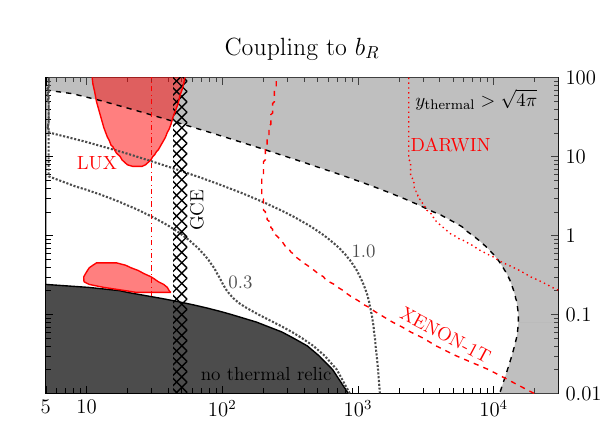}

\hspace{-0.7cm}
\includegraphics[scale=1.05]{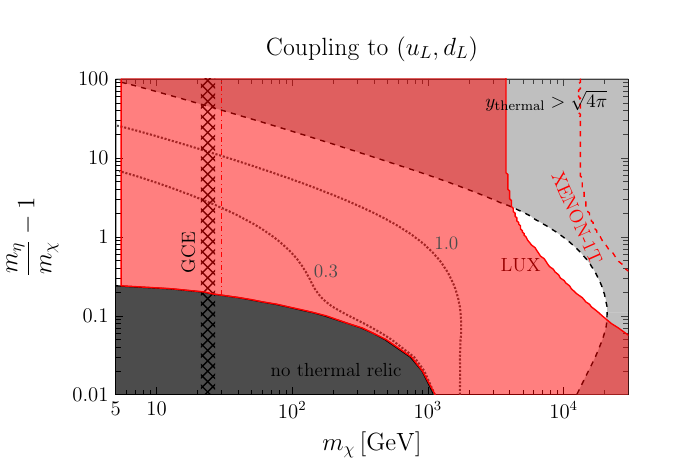}
\hspace{-1.0cm}
\includegraphics[scale=1.05]
{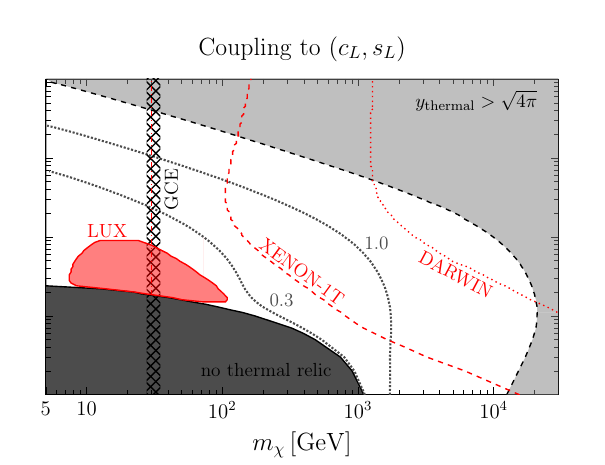}
\hspace{-1.0cm}
\includegraphics[scale=1.05]{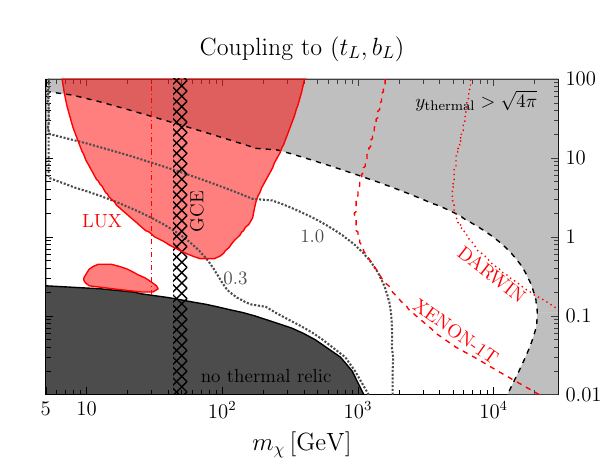}
\caption{\small Regions of the parameter space for Dirac dark matter particles with scalar mediator excluded by the LUX experiment, assuming coupling to right-handed up-quarks (upper panels), right-handed down-quarks (middle panels), and left-handed quark doublets (lower panels), and assuming that the dark matter density of the Universe is generated by thermal freeze-out (the necessary value of the Yukawa coupling is indicated by the dotted black contour lines). The dark gray shaded regions in the lower left corners of each figure are theoretically inaccessible, due to efficient coannihilations which deplete the dark matter density below the observed value, while the light gray shaded regions violate our perturbativity condition $y_\text{thermal} < \sqrt{4 \pi}$. }
\label{fig:thermal_limits_quarks}
\end{figure}

\begin{figure}[t!]
\hspace{-0.7cm}
\includegraphics[scale=1.05]{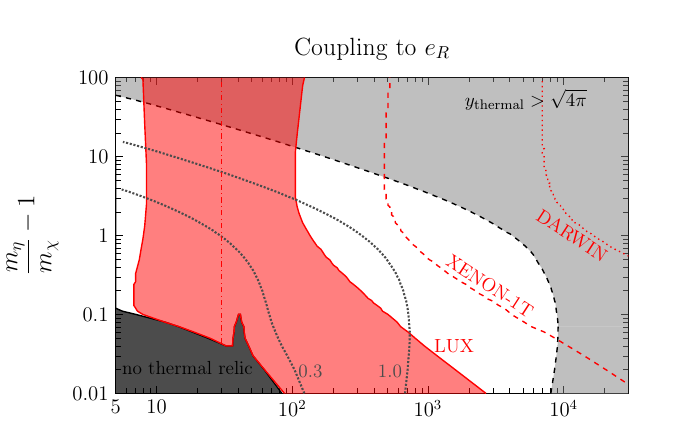}
\hspace{-1.0cm}
\includegraphics[scale=1.05]
{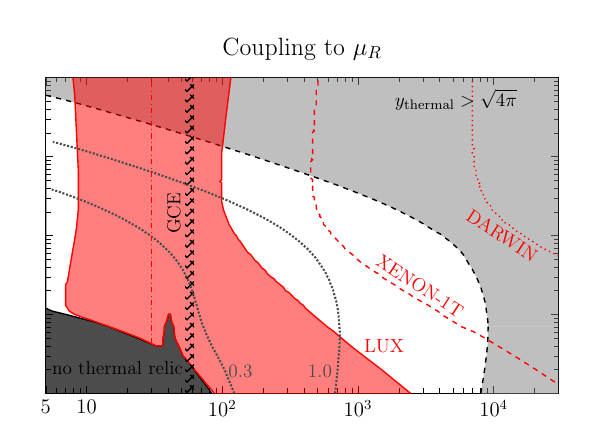}
\hspace{-1.0cm}
\includegraphics[scale=1.05]{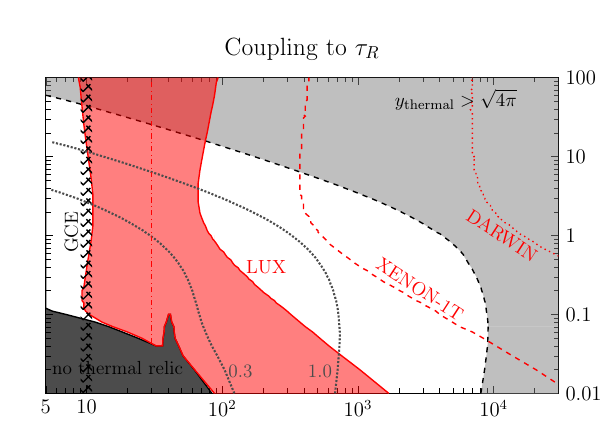}

\hspace{-0.7cm}
\includegraphics[scale=1.05]{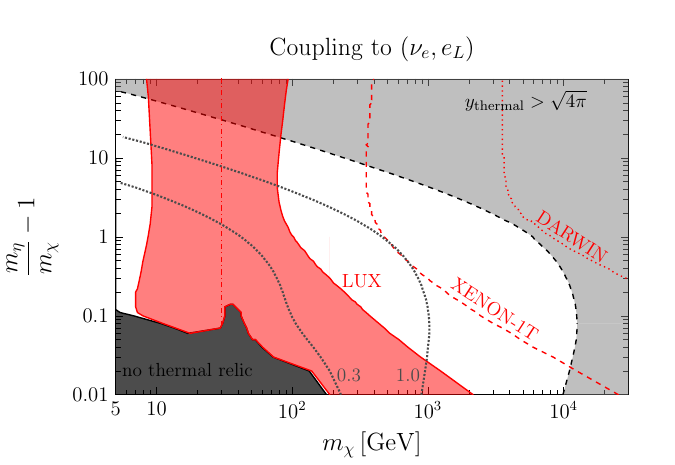}
\hspace{-1.0cm}
\includegraphics[scale=1.05]{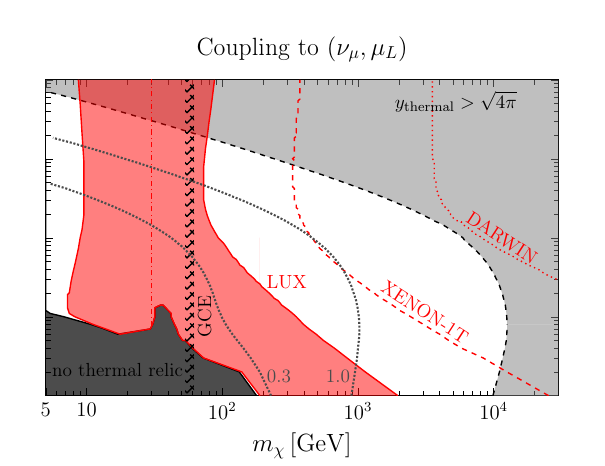}
\hspace{-1.0cm}
\includegraphics[scale=1.05]{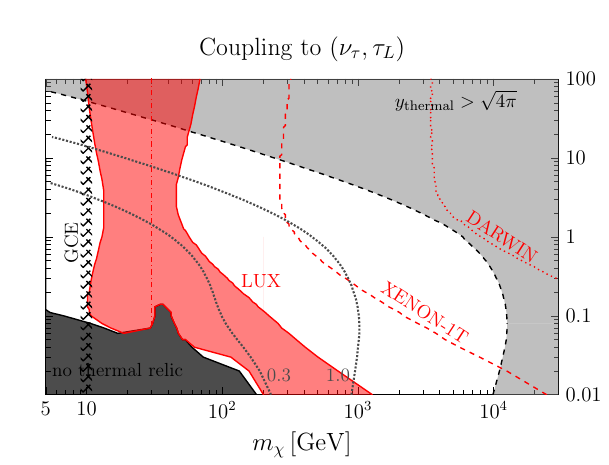}
\caption{\small Same as Fig.~\ref{fig:thermal_limits_quarks}, but for dark matter coupling to right-handed charged leptons (upper panels) and to left-handed lepton-doublets (lower panels).}
\label{fig:thermal_limits_leptons}
\end{figure}

We finally determine the regions of the parameter space which are excluded by the LUX experiment for each of these scenarios, as well as the regions that can be probed with the future XENON1T and DARWIN experiments, fixing the value of the Yukawa coupling at each point to the value $y_\text{thermal}$ required to reproduce the observed dark matter abundance via thermal freeze-out. The results for the case of dark matter coupling to quarks are shown in Fig.~\ref{fig:thermal_limits_quarks}, and for coupling to leptons in Fig.~\ref{fig:thermal_limits_leptons}. The dotted black contour lines show the value of the Yukawa coupling which leads, at that point of the parameter space, to the observed dark matter abundance $\Omega_{\rm DM} h^2\simeq 0.12$. Besides, for each of the possible scenarios, the region of the parameter space corresponding to low mass and small mass splitting is theoretically inaccessible since the coannihilations suppress the relic abundance of $\chi$ below the observed value. This region is shown in Figs.~\ref{fig:thermal_limits_quarks} and ~\ref{fig:thermal_limits_leptons} as dark grey. On the other hand, large masses and/or large mass splittings correspond to Yukawa couplings $y>\sqrt{4\pi}$, where our perturbative calculation loses validity.\footnote{This condition is motivated by the requirement that  the expansion parameter $\alpha_y \equiv y^2/4 \pi$ for higher-order corrections should be smaller than one.} It is worthwhile noticing that this requirement sets an upper limit on the dark matter mass which approximately reads $m_{\rm DM}\lesssim 20$ TeV for coupling to quarks and $m_{\rm DM}\lesssim 15 $ TeV for coupling to leptons.

The left panels in Fig.~\ref{fig:thermal_limits_quarks} correspond to scenarios where $\chi$ couples to first generation quarks, namely to right-handed up-quarks (top panel), right-handed down-quarks (central panel), and to the first generation left-handed quark doublet (lower panel). As apparent from the plots, practically the whole parameter space for a thermal relic compatible with our perturbativity condition is already excluded by the LUX experiment, with XENON1T being able to constrain dark matter masses up to $\simeq 10$ TeV, thereby probing the whole parameter space of the model. These strong direct detection bounds result from the unsuppressed tree-level coupling of the dark matter particle $\chi$ to light quarks, \textit{cf.} Eq.~(\ref{eq:Leff_tree}).

On the other hand, for coupling to quarks of the second generation, the tree-level coupling is negligible and the limits dominantly arise from the dark matter electromagnetic moments, \textit{cf.} Fig.~\ref{fig:MostImportantContris_QuarkCoupling}. Consequently, the constraints from LUX are considerably  weaker, as apparent from the second column of Fig.~\ref{fig:thermal_limits_quarks} for dark matter coupling to $c_R$ (top panel), $s_R$ (central panel), and $(c_L, s_L)$ (lower panel). Nevertheless, it is worth pointing out that despite the suppression of the one-loop diagrams, LUX already excludes dark matter masses up to $\simeq 100-500$ GeV, depending on the mass splitting of $\chi$ and $\eta$ as well as on the precise coupling scheme of dark matter. Most interestingly, in these scenarios there are exciting prospects for the upcoming direct detection experiments to probe large parts of the viable parameter space. For example, a DARWIN-like experiment would be able to cover practically the whole parameter space of thermally produced dark matter, compatible with our perturbativity condition, despite the fact that all direct detection processes occur only at the one-loop level. Furthermore it is worthwhile remarking that direct detection experiments have an enhanced sensitivity in the case of mass-degenerate scenarios, i.e. $m_\eta/m_\chi \simeq 1$, as apparent from the figures. This larger sensitivity follows from the parametric dependence of the charge radius operator $ b_\chi \propto 1/( (m_\eta/m_\chi)^2 - 1)$, which leads to an enhancement of the scattering cross section as the mass spectrum becomes more and more degenerate.

Lastly, for dark matter coupling to third-generation quarks, the constraints on the parameter space are presented in the right panels of Fig.~\ref{fig:thermal_limits_quarks}. In the case of coupling to $t_R$, shown in the upper right panel, we restrict our analysis to $m_\chi > m_t \simeq 175$ GeV, which ensures that the process $\chi \bar{\chi} \rightarrow t \bar{t}$ is kinematically allowed during freeze-out in all of the considered parameter space. In principle, a thermal relic can also exist for $m_\chi < m_t$, however in that case the relic density would instead be set by one-loop annihilations or two-to-three processes, changing the phenomenology of the model quite dramatically. For an extensive discussion of the model in that part of the parameter space we refer to~\cite{Kumar:2013hfa}. As argued in section \ref{sec:L_eff}, if the dark matter particle couples to either $t_R$ or $(t_L, b_L)$, the scattering rate in direct detection experiments is typically dominated by the $Z$ exchange contribution, while it is dominated by the electromagnetic moments for coupling to the right-handed bottom quark. Then, for coupling to the right-handed top quark, LUX excludes masses up to $200 - 400$ GeV, depending on the mass splitting of $\eta$ and $\chi$, while the constraints for coupling to $b_R$ (central panel) are somewhat weaker due to the absence of a large effective dark matter coupling to the $Z$ boson. Again, it is important to point out that XENON1T and eventually DARWIN will probe a significantly larger region of the parameter space than LUX, and will be able to rule out, or hopefully discover, dark matter particles with masses up to several TeV. Similar conclusions also hold for the case of $\chi$ coupling to the third generation quark-doublet, shown in the lower right panel of Fig.~\ref{fig:thermal_limits_quarks}.

Direct search experiments are subject to astrophysical and nuclear uncertainties which might modify the conclusions presented above. Hence, and in order to robustly exclude a region of the parameter space or to confirm a possible signal, it is important to exploit the complementarity with indirect and collider searches, which rely on very different physical processes, and which are affected by different systematic uncertainties. A thorough investigation of the complementarity among the different search strategies for Dirac dark matter is beyond the scope of this paper and here we will limit ourselves to briefly comment on the qualitative features of the indirect and collider searches. If the colored mediator is sufficiently light, it might be directly produced at the LHC via strong interactions. If, moreover, the mass difference between the scalar mediator and the dark matter particle is sizable, the subsequent decay produces a characteristic signal consisting in two jets plus missing transverse momentum, which has been searched for in the framework of simplified supersymmetric models \cite{Chatrchyan:2014lfa,Aad:2014wea,Aad:2014vma,Khachatryan:2014qwa}. In contrast, when the mass difference is small, the jets might be too soft to be detected and a different search strategy should be pursued, for instance the search for monophoton~\cite{Aad:2014tda} or monojet events~\cite{Aad:2014nra} plus missing transverse momentum. Finally, for very heavy mediators, only the dark matter particle can be pair-produced at the collider. In this regime, dark matter signals could be detected in searches for monojet, monophoton or mono-W/Z boson plus missing transverse momentum~\cite{Aad:2015zva,atlasMono,Khachatryan:2014rra}. Present collider limits constrain some regions of the parameter space with $m_\chi\lesssim 200$ GeV, although the precise reach of the limits crucially depends on the details of the model. Additional constraints from colliders follow from electroweak precision observables, notably the $Z$-boson decay width~\cite{ALEPH:2005ab}, and which exclude the existence of scalar mediators with mass $m_\eta\lesssim 40$ GeV. 

On the other hand, the strongest indirect detection limits on this class of scenarios stem from the non-observation of an excess of gamma-rays correlated to the direction of dwarf galaxies. A recent analysis by the Fermi-LAT collaboration from a stacked analysis of 15 dwarf spheroidal galaxies excludes $m_{\rm DM}\lesssim 100$ GeV for dark matter coupling to the $b$-quark~\cite{Ackermann:2015zua}, which is competitive, and for some choices of parameters better, than the LUX limits shown in Fig.~\ref{fig:thermal_limits_quarks}. Additional constraints on the model can be derived from gamma-ray observations of the Milky Way center, although this search is affected by large, and still not fully understood, backgrounds. Interestingly, some authors have claimed evidence for a gamma-ray excess in this region~\cite{Goodenough:2009gk,Hooper:2010mq,Hooper:2011ti,Abazajian:2012pn,Daylan:2014rsa,Calore:2014xka}, which can be interpreted in terms of dark matter annihilations. Dirac dark matter with a charged mediator constitutes an excellent particle physics framework to explain the excess (in contrast to Majorana or scalar dark matter, which have annihilation cross sections at the Galactic center which are helicity- and velocity-suppressed). The mass ranges favored by the dark matter interpretation of the galactic center excess (GCE)  for the annihilation final states selected in~\cite{Calore:2014nla}, and which point to an annihilation cross section of the order of the thermal value, are shown in Fig.~\ref{fig:thermal_limits_quarks}, as vertical hatched bands. It follows from the figure that direct detection experiment can test, for some channels, this interpretation of the excess, especially for couplings to first generation quarks, to $s_R$ or to $c_R$. One should bear in mind that the astrophysical uncertainties in the extraction of the dark matter parameters from the GeV excess are still large. Therefore, the value of the annihilation cross section can be significantly lower than the thermal value and accordingly the impact of the LUX limits.

Finally, we turn to scenarios in which the dark matter particle only couples to one of the right- or left-handed leptons of the Standard Model. As discussed in Section~\ref{sec:L_eff}, in this scenario, the tree-level interactions vanish, however the one-loop diagrams induce significant scattering rates in direct detection experiments. In fact, many of the qualitative features of the scenario where $\chi$ couples to heavy quarks also apply to this case. The parameter space for the various Dirac dark matter scenarios with coupling to leptons are shown in Fig.~\ref{fig:thermal_limits_leptons}, where again the region of the parameter space incompatible with thermal production and perturbativity are shown, respectively, as dark and light gray shaded regions. For coupling to leptons, the region of parameter space for which no thermal relic exists due to strong coannihilations  is smaller compared  to the case of dark matter coupling to quarks, since  for leptophilic dark matter the relevant coannihilation channels (such as $\eta \eta^\dagger \rightarrow \gamma \gamma$) are suppressed by powers of the electromagnetic coupling, while for coupling to quarks, the corresponding processes are induced by the strong coupling between the scalar $\eta$ and the gluons. Besides, in the case of dark matter coupling to leptons, for $m_\chi \simeq m_\eta \simeq 45$ GeV the coannihilation process $\eta \eta^\dagger \rightarrow f \bar{f}$ is resonantly enhanced due to the diagram where the $Z$ boson is exchanged in the $s$-channel. Hence, around that dark matter mass, the region in parameter space where no thermal relic can exist due to coannihilations is extended to higher mass splittings.

The region of parameter space excluded by LUX, as well as the range of masses which will be accessible to XENON1T and DARWIN are very similar for the six different scenarios shown in Fig.~\ref{fig:thermal_limits_leptons}. For all mass splittings between the mediator $\eta$ and the dark matter particle $\chi$, LUX excludes dark matter masses between $8 \text{ GeV} \lesssim m_\chi \lesssim 100$ GeV, and XENON1T will be sensitive up to $m_\chi \simeq 500$ GeV. Moreover, as noted already before, the charge radius operator $b_\chi$ is enhanced for small mass splittings, leading to an increased sensitivity of the direct detection experiments for $m_\eta/m_\chi \rightarrow 1$. Moreover, from these figures it follows that a multi-ton detection experiment like DARWIN will practically cover the whole parameter space of these scenarios, again despite the fact that all direct detection signals are one-loop suppressed. 

Scenarios with Dirac dark matter coupling to leptons can also be probed in collider or indirect search experiments, which give complementary limits on the parameter space of the model. The charged scalar mediator can be pair-produced via the Drell-Yan process in colliders and subsequently decay producing a characteristic signature of opposite-sign, same-flavor leptons, plus missing transverse momentum. This search has been undertaken in the context of slepton searches at the LHC by the ATLAS~\cite{Aad:2014vma} and CMS collaborations~\cite{Khachatryan:2014qwa}, and at LEP II by the ALEPH, DELPHI, L3 and OPAL collaborations~\cite{LEP-sleptons}, and exclude some regions of the parameter space with $m_\chi\lesssim 100$ GeV (see \cite{Garny:2015wea} for a compilation of the limits of the parameter space of a Majorana dark matter particle that couples to a lepton and a scalar mediator, and which can  also be applied to Dirac dark matter, since the production mechanism of scalar mediators and their decay modes are identical in both cases). Finally, the precise measurement of the $Z$-boson decay width at LEP~\cite{ALEPH:2005ab} excludes mediators with mass $m_\eta\lesssim 40$ GeV.

Stringent limits on this scenario also follow from indirect searches. The annihilation into leptons produces an additional component in the cosmic positron flux which is severely constrained by the AMS-02 data on the positron fraction~\cite{Aguilar:2013qda,Accardo:2014lma}, defined as the flux of positrons divided by the total flux of electrons plus positrons, as well as the data on the positron flux~\cite{Aguilar:2014mma}. More specifically, the non-observation of the sharp feature in the positron fraction produced by the hard leptons produced in these annihilation channels translates into strong limits on the parameters of the model. From the limits on the cross section derived in~\cite{Bergstrom:2013jra} one can exclude $m_{\chi}\lesssim 120$ GeV for coupling to $e_R$, $m_{\chi}\lesssim 70$ GeV for coupling to $\mu_R$ and $m_{\chi}\lesssim 30$ GeV for coupling to $\tau_R$. The more conservative limits from~\cite{Ibarra:2013zia} exclude $m_{\chi}\lesssim 100$ GeV and $m_{\chi}\lesssim 40$ GeV for couplings to $e_R$ and $\mu_R$, respectively. Strong constraints on these channels can also be derived from dwarf galaxy observations, which exclude $m_\chi\lesssim 100$ GeV~\cite{Ackermann:2015zua}.

\section{Conclusions}
\label{sec:conclusions}

In some scenarios, the dark matter particle only couples to the nucleons via quantum effects, with a scattering rate which is expected to be very small. On the other hand, the excellent sensitivity of current direct search experiments might allow to identify such rare events over the environmental and instrumental backgrounds, thus opening the possibility of probing a class of models which were not accessible to previous experiments, such as those where the dark matter couples only to leptons or to heavy quarks.

In this paper, we have investigated this exciting possibility in the framework of a toy model where the dark matter is a Dirac fermion, singlet under the Standard Model gauge group, which couples to one of the Standard Model fermions via a Yukawa coupling with a scalar mediator. For reasonable values of the parameters of the model, the observed dark matter abundance $\Omega_{\rm DM} h^2\simeq 0.12$ can be explained via thermal freeze-out. In this model, the scattering rate with nucleons receives various contributions, which depend on the choice of the  Standard Model fermion. For couplings to first generation quarks, the dark matter interacts with the nucleons at tree level  resulting in a comparatively large scattering rate. On the other hand, for couplings to any  Standard Model fermion, the dark matter interacts with the nucleon via one-loop penguin diagrams mediated by the photon, the $Z$ boson or the Higgs boson. Lastly, for couplings to any colored fermion, the dark matter interacts with the nucleon also via one-loop box diagrams with two external gluons.

Remarkably, current experiments have reached the necessary sensitivity to probe these quantum-induced dark matter interactions with nucleons. More specifically, assuming thermal production for the dark matter population in our Universe, the current LUX data probe the parameter space of {\it all} scenarios considered in this paper, posing in some cases very stringent constraints on the parameter space of the model. For instance,   the LUX data exclude dark matter masses below $\sim 3$ TeV for coupling to first generation quarks, masses between $\sim 10-100$ GeV for coupling to the right-handed charm or strange quarks, and masses also between $\sim 10-100$ GeV for coupling to leptons (of any generation and chirality). The future XENON1T experiment will continue closing in on the parameter space of the model and will be able to probe dark matter masses up to several hundreds of GeV, while in the longer term, a multi-ton xenon detector such as DARWIN will be able to exclude masses as large as a few TeV. Direct search experiments will then have a high sensitivity to this class of scenarios, even for multi-TeV dark matter masses, which are difficult to probe at the LHC. 

\vspace{0.5cm}
\section*{Acknowledgements}
We are grateful to Riccardo Catena for useful discussions. This work was partially supported by the DFG cluster of excellence ``Origin and Structure of the Universe,'' the TUM Graduate School and the Studienstiftung des Deutschen Volkes.

\vspace{0.5cm}
\appendix

\section{Derivation of the limits from the LUX data}
\label{sec:appendix_lux_limits}
 We recall from Section \ref{sec:event_rate} that the limits on the spin-independent scattering cross sections published by the LUX collaboration cannot be applied to our scenario, due to the large isospin violation and, especially, due to the different dependence of the differential cross section with the recoil energy, as a result of the scatterings mediated by a photon. In this appendix we briefly describe our method to translate the LUX null results into limits on the Dirac dark matter model considered in this paper. Starting from the  differential dark matter-nucleus scattering cross section, Eq.~(\ref{eq:scattering_crosssection}), the number of expected recoils per kilogram target mass and per keV recoil energy can be straightforwardly calculated, the result being:
\begin{align}
\frac{\text{d}R_i}{\text{d} E_R} = \frac{\xi_i \rho_{\odot}}{m_{N_i} m_\chi} \int_{v_\text{min}(E_R)}^{\infty} \text{d}v \, v f_E(v) \frac{\text{d}\sigma}{\text{d}E_R} \quad \text{with} \quad v_\text{min}(E_R) = \sqrt{\frac{m_{N_i} E_R}{2 \mu_\text{red}^2}} \,.
\label{eq:diff_scattering_rate}
\end{align}
In this expression, the index $i$ refers to a given xenon isotope with mass $m_{N_i}$ and mass fraction $\xi_i$, $\rho_\odot \simeq 0.3 \text{ GeV}/\text{cm}^3$ is the local dark matter density (of which $\rho_\odot/2$ is made up by $\chi$ and $\rho_\odot/2$ by $\bar \chi$), and $f_E(v)$ is the one-dimensional speed distribution of dark matter particles in the Earth's rest frame. The latter is derived from an isotropic Maxwell-Boltzmann distribution in the Galactic rest frame with velocity dispersion $v_0=220$ km/s, using a mean Earth's velocity of $v_E=233.3$ km/s.

The total expected scattering rate in the LUX reads~\cite{DelNobile:2013sia}
\begin{align}
R = \frac{1}{2} \int_{S_1^{\text{min}}}^{S_1^{\text{max}}} \text{d} S_1 \, \epsilon \left(S_1\right) \sum_{n=1}^{\infty} \text{Gauss}\left( S_1 | n, \sqrt{n} \, \sigma_{\text{PMT}} \right) \int_{E_R^{\text{min}}}^\infty \text{d}E_R \, \text{Poiss} \left( n | \nu(E_R) \right) \sum_i \frac{\text{d}R_i}{\text{d} E_R}
\label{eq:total_scattering_rate}
\end{align}
Here $\nu(E_R)$ is the expected number of photoelectrons (PE) for a nuclear recoil with energy $E_R$ which we extract from Fig.~4 of~\cite{Akerib:2013tjd} by taking the intersections of the red solid line with the contours of constant $E_R$. The actual number of photoelectrons $n$ is then given by a Poisson distribution with mean $\nu(E_R)$; following the LUX collaboration, we conservatively do not model any recoils below $E_R^\text{min}=$ 3 keV. We take into account the single-photoelectron resolution of the photomultipliers by assuming that the observed $S_1$ signal is given by a Gauss distribution with mean $n$ and standard deviation $\sqrt{n} \, \sigma_{\text{PMT}}$, for which we take $\sigma_{\text{PMT}}=0.37$~\cite{Akerib:2012ys}. The acceptance $\epsilon \left( S_1 \right)$ of the LUX detector is given by the red dashed line in Fig.~1 of~\cite{Akerib:2013tjd}, and the additional factor $1/2$ in Eq.~(\ref{eq:total_scattering_rate}) takes into account that we define only the region below the mean nuclear recoil band (red solid line in Fig.~4 of~\cite{Akerib:2013tjd}) as the signal region for dark matter recoils; see~\cite{DelNobile:2013gba} for an in-depth discussion of this choice. Finally, $S_1^{\text{min}}= 2$ PE and $S_1^{\text{max}}= 30$ PE are the minimal and maximal number of photoelectrons considered in the LUX analysis.

In the signal region, LUX observed one event after running with an exposure of $85.3 \times 118.3 \text{ kg}\cdot\text{days}$. Making no assumptions about the number of expected background events, we obtain a $90 \%$ C.L. upper limit of $N_\text{max}=3.89$ expected recoil events, which then can be used to set limits on the normalization of the dark matter scattering cross section. As a cross-check we compare our limits on an isospin-conserving spin-independent interaction with those published by the LUX collaboration~\cite{Akerib:2013tjd}: depending on the dark matter mass, our limits are more conservative by a factor $1.5 - 2.5$.

\bibliographystyle{JHEP-mod}
\bibliography{references}

\end{document}